\documentclass{article}

\usepackage{graphicx,hyperref}

\title{Complementarity of generative models for road networks}
\author{J. Raimbault$^{1,2,3,\ast}$\\
$^{1}$ Center for Advanced Spatial Analysis, University College London\\
$^{2}$ UPS CNRS 3611 ISC-PIF\\
$^{3}$ UMR CNRS 8504 G{\'e}ographie-cit{\'e}s\medskip\\
$\ast$ \texttt{juste.raimbault@polytechnique.edu}
}
\date{}

\begin{document}

\maketitle

\begin{abstract}
	Understanding the dynamics of road networks has theoretical implications for urban science and practical applications for sustainable long-term planning. Various generative models to explain road network growth have been introduced in the literature. We propose in this paper a systematic benchmark of such models integrating different paradigms (spatial interactions, cost-benefit compromises, biological network growth), focusing on the feasible space of generated network measures. We find a quantitatively high complementarity between the different models. This confirms the necessity of a plurality of urban models, in line with integrative approaches to urban systems.
\end{abstract}

\section{Introduction}

Territorial systems are complex in the sense that they span across multiple dimensions, and are produced by various agents at different spatial and temporal scales \cite{batty2007complexity}. They are furthermore adaptive and self-organised, as witnesses the resilience of co-evolving city systems on long time scales \cite{pumain2021co}. Among the built environment, environmental, and social components, one crucial driver of spatial structure and dynamics of these territorial systems are transportation infrastructure networks, and more particularly road networks. They irrigate territories by distributing accessibility, and their topology and hierarchy often shapes spatial interactions and thus urban and land-use dynamics \cite{wegener2004land}. Strong effects of path-dependance furthermore highlights the role of these networks \cite{ribeill1985aspects}.

Although accessibility is nowadays strongly multimodal \cite{cats2021multi}, road networks have been the dominant transportation infrastructure in urban systems over long historical periods \cite{verdier2007extension}, and they still are an important aspect of both urban form at the microscopic scale and accessibility at the mesoscopic and macroscopic scales. Understanding the link between underlying processes driving the growth of road networks and their shape is an important subject, towards potential application to sustainable planning.

Generative processes for road networks are multiple: for example a combination of self-organisation and top-down planning may leave a significant signature in topological features of these networks \cite{barthelemy2013self}. In order to explain such evolution and in some cases reproduce existing networks, multiple simulation models have been introduced in the literature by different disciplines. These span across a broad spectrum, from data-driven models to parsimonious stylised models. \cite{xie2009modeling} propose a broad survey of such models, including economics, transport geography, transport planning and network science. The diversity of processes taken into account can range from economics of investment \cite{xie2011evolving}, to negotiations between planning stakeholders \cite{raimbault2021introducing}, full top-down planning \cite{szell2021growing}, self-organisation through morphogenesis \cite{tirico2018morphogenesis}, or local optimisation \cite{barthelemy2009co}, among others.

Most of these models include empirically documented processes, and have been validated against stylised facts such as hierarchical organisation of the network or the distribution of network measures, and also often compared against existing networks with reasonable success. However, to what extent these processes and models are complementary remains an open question. We propose in this paper a systematic benchmark of several road network growth models, to clarify the necessity of considering such diverse models. We extend the multi-modeling approach of \cite{raimbault2018multi} by focusing on the feasible space of network measures as model outputs. We consider in particular the following mechanisms for link addition, which were chosen to illustrate a diversity of underlying assumptions and frameworks: (i) gravity potential deterministic breakdown \cite{raimbault2019second}; (ii) random potential breakdown \cite{raimbault2020unveiling}; (iii) cost-benefits compromise  \cite{louf2013emergence}; (iv) biological network generation  \cite{raimbault2018systemes}.

The rest of this paper is organised as follows: we first describe in details each road network growth model included in the benchmark; we then describe the implementation and model setup, followed by numerical experiments aimed at comparing feasible spaces for each model; we finally discuss implications and possible extensions.

\section{Road network growth models}

\subsection{Baseline dynamics}

All models are included within the same baseline dynamics, for a better comparability. We use the co-evolution model between population density and road network described by \cite{raimbault2019urban}, but evolve the network only. At a mesoscopic spatial scale, given a population density distributed as a grid (typically with size 50x50), road network nodes and links are added iteratively. More precisely, starting from an empty network, at each time step:

\begin{enumerate}
	\item new network nodes are added, with a preferential attachment to population and to existing roads to determine the cell, and a random position within the cell; a fixed number $N_n = 20$ of nodes is considered;
	\item added nodes are connected to the existing network with a connectivity algorithm: the closest network component is determined, and a shortest distance connection is created (either to the closest node, or with a perpendicular projection to the closest link);
	\item a fixed number of links $N_l$ (model parameter) is added, following one of the models included in the benchmark;
	\item the resulting network is made planar, by adding nodes at intersections of crossing links. 
\end{enumerate}

The model can be run for a fixed number of time steps $t_f$, or until a total number of links $N_f$ is reached.

\subsection{Link addition models}

\paragraph{Null model}

The null model corresponds to no additional link besides the ones added to keep the network connected. This yields tree networks with no loops.

\paragraph{Random links}

A second null model consists in the addition of random links. A random node is chosen, and connects to an other node to which it is not already connected. This procedure is done $N_l$ times.

\paragraph{Random gravity potential breakdown}

The model explored in \cite{raimbault2020unveiling} adds random links according to spatial interactions between nodes. Given a population for each node $P_i$ (determined here by the sum of the population within the catchment area of each node), a classical gravity model gives interaction forces between nodes. The origin city of the new link is chosen hierarchically to population, and the destination city following spatial interactions. The link is realised if the current network imposes a high detour (i.e. if the rate between euclidian distance and network distance is below a given threshold). This procedure is repeated until $N_l$ links have been added.

\paragraph{Deterministic gravity potential breakdown}

The model formulated by \cite{raimbault2019second} also add links by considering spatial interactions between nodes. Still accounting for node populations, a spatial interaction model yields interaction forces between nodes (see \cite{raimbault2019second} for detailed equations). Potential links are taken as a fixed proportion among all couples, with the highest potential. A cost criteria at the last stage selects the links with the lowest detour among the potential links. Note that an other heuristic assuming strong top-down planning and investment could at this stage select links with the highest detour.

\paragraph{Cost-benefit aggregation}

The cost and benefits included in the previous heuristic are also considered through aggregation by \cite{louf2013emergence}. For each node couple, a cost function aggregating interaction potential with construction cost (as the constructed link distance) normalised by an aggregation parameter $\lambda$, is computed. The $N_l$ links with the lowest cost are constructed.

\paragraph{Biological network generation}

\cite{raimbault2018systemes} uses a slime-mould model to generate synthetic networks. We use the same model with the following process: (i) generate a background grid of local links between neighbour cells, on top of the existing network; (ii) run the slime-mould self-reinforcement model in this provisory network with origins/destinations as node couples with population, for a given number of steps $t_b$; (iii) remove all provisory link with a capacity below a fixed threshold $\theta_b$; (iv) simplify the resulting network (by removing nodes with degree exactly two) and keep the new links to add them to the evolving network.

\subsection{Indicators}

The generated road networks are quantified using the following measures: diameter, mean path length, mean betweenness centrality, mean closeness centrality, mean detour, cumulated length. More indicators could be taken into account but \cite{raimbault2019urban} showed that these already provided a reasonable coverage.

\section{Results}
 
\subsection{Data}
  
We setup the model with real population grids. Therefore, the Eurostat population grid was used \cite{batista2013high}. 50 configurations were randomly selected, equally distributed into morphological clusters provided by \cite{raimbault2018multi} (measures of the spatial distribution of population). The initial population distribution is considered as a model parameter.
  
\subsection{Implementation} 

The models are implemented into a single one in the NetLogo software, and integrated into the OpenMOLE software for model exploration and validation \cite{reuillon2013openmole}. The models are open source and available at \url{https://github.com/JusteRaimbault/NetworkGrowth}


\subsection{Exploration}

We explore the parameter space of each model by using a Latin Hypercube Sampling, and sampling 40,000 parameter points for each model. The index of population configuration is included as a parameter in each case. As each model has a low number of parameters (below 5, for the deterministic gravity breakdown), we assume that the obtained point cloud cover most of the feasible space. More robust results should be obtained with diversity search algorithms as detailed in the discussion below.

\begin{figure}
\centering
\includegraphics[width=\linewidth]{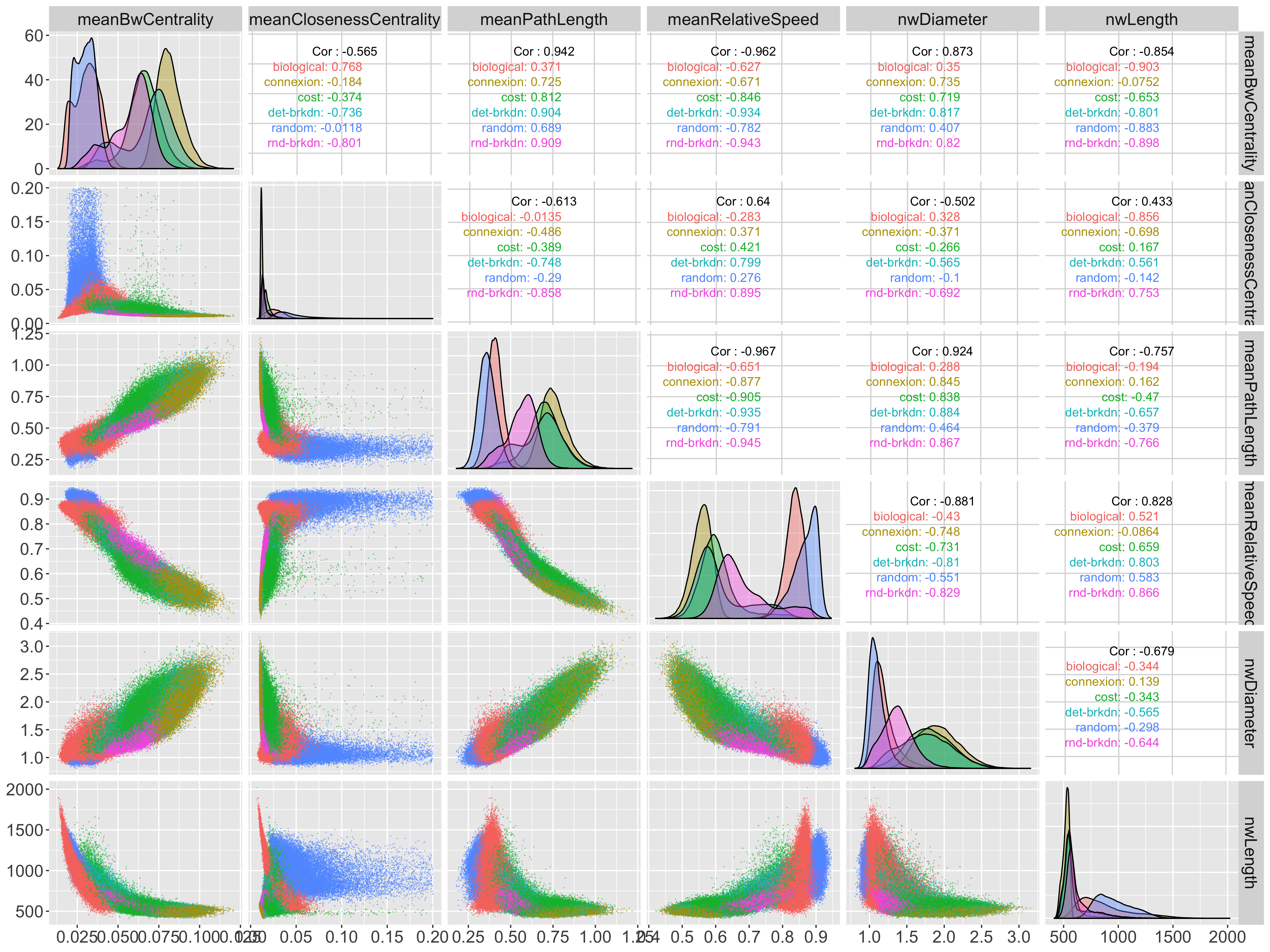}
\caption{Scatterplot of feasible space for each model, between all indicators (rows and columns). The diagonal histograms give the statistical distribution of each indicators. Color give the model. Estimated correlations between indicators, stratified by model, are also given.\label{fig:scatter}}
\end{figure}

We show in Fig.~\ref{fig:scatter} the scatterplot of indicators across models. We obtain overall important parts of the space which are occupied by a single model, suggesting a complementarity. Surprisingly, the closest model to the random null model seem to be the biological network, while the connection heuristic is close to the cost-benefit (which could be expected as only local links are added in the latest case). The random breakdown seems to be the model the most distinguished from others. We also find different performance patterns, as in terms of relative speed, connection model has the lowest while the random has the highest (this indicator is not weighted by population).

\begin{figure}
\centering
\includegraphics[width=\linewidth]{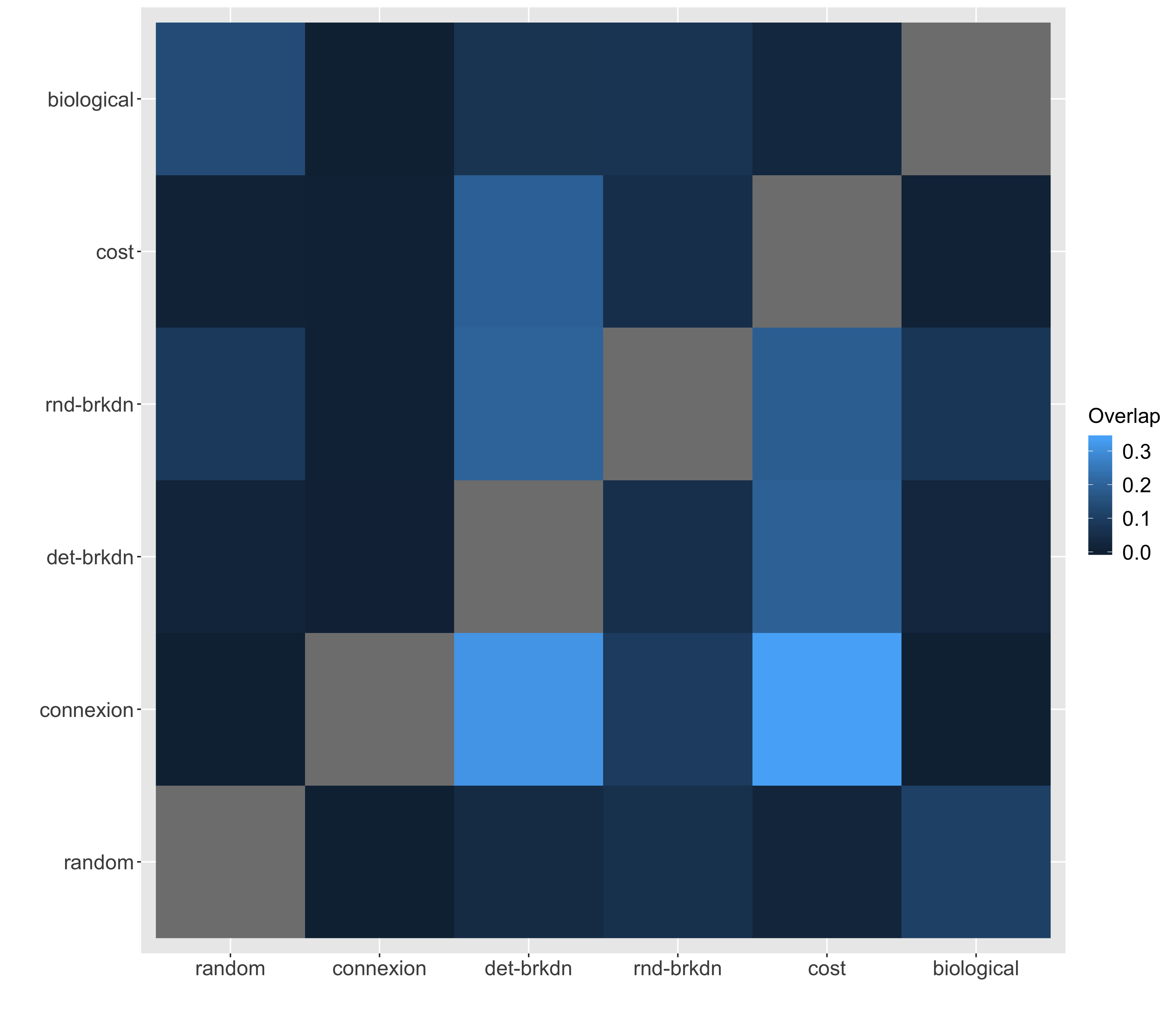}
\caption{Estimated overlaps between hypervolumes of model outputs.\label{fig:overlap}}
\end{figure}

To further quantify the complementarity between models, we compute the overlap between point clouds. Therefore, we estimate hypervolumes and compute their relative intersection as $I(O1 \rightarrow O2) = Vol(O1 \cap O2) / Vol(O2)$. This non-symmetric measure gives the proportion of point cloud $O2$ overlapping with $O1$. The corresponding matrix is shown in Fig.~\ref{fig:overlap}. We find a relatively low level of overlapping, as the highest is just above 0.3, for the connection cloud in the deterministic breakdown and the cost. More than half of couples are below 0.1, confirming that point clouds are mostly distinct in the multidimensional indicator space.

Overall, this numerical exploration confirms the complementarity of the diverse heuristics included in our benchmark.

\section{Discussion}

Our result contribute to a general effort of urban models comparison and benchmarking, which is an essential step towards the construction of integrative models and theories. We have shown quantitatively a strong complementarity between diverse processes driving the growth of road networks. This confirms the necessity to include a plurality of models, views and disciplines in the understanding of road network dynamics.

Current work is focused on a more systematic determination of feasible spaces. To that end, we use the Pattern Space Exploration (PSE) algorithm introduced by \cite{cherel2015beyond} and implemented into the OpenMOLE platform. This algorithm, which is a particular case of diversity search algorithms, has been shown to be efficient to discover a large number of patterns and effectively cover output space in low dimensions. This would allow obtaining more robust measures of the complementarity of the different models included.

An other important ongoing extension is the calibration of models on real network measures. In this work, real setups were used only for population distributions. Using databases with a broader span such as the GHSL database \cite{florczyk2019ghsl}, more urban regions could be considered. Corresponding street networks could then be extracted from OpenStreetMap as it was done in \cite{raimbault2019urban}, and real network measures included in the point cloud. Depending on the obtained coverage, specific calibration algorithms may be needed to try approaching the points out of the simulated space.

Finally, this work would gain to include more generative models for road networks. \cite{molinero2020model} recently proposed a network model based on self-reinforcement processes. \cite{queyroi2018biological} compares biological with shortest-path routing procedures to simulate transport networks. \cite{mimeur:tel-01451164} introduces a model for rail networks which includes the progressive hierarchisation of the network. The governance processes core to the model developed by \cite{raimbault2021introducing} or the investment processes described in \cite{cats2020modelling}, allow inferring results on other dimensions of the transport systems, but it is not clear to what extent they are able to produce fundamentally different networks than the ones obtained through simpler models. All these other approaches, among others, are also relevant candidates to be included in such systematic benchmark. One difficulty remains to be able to formalise, setup and implement all models within a single comparable context.

\end{document}